\documentclass{iopjournal}
\usepackage{graphicx}
\usepackage{float} 
\usepackage{amsmath}
\usepackage{dcolumn}
\usepackage{bm}

\usepackage[utf8]{inputenc}
\usepackage[T1]{fontenc}
\usepackage{mathptmx}
\usepackage{etoolbox}
\usepackage[separate-uncertainty]{siunitx}
\usepackage{comment}
\usepackage{lmodern}

\usepackage[numbers,sort&compress]{natbib}

\usepackage{printlen}
\uselengthunit{cm}

\begin{document}


\title{Radiation safety considerations for ultrafast lasers beyond laser machining}
\author{Simon Bohlen$^{1,*}$\orcid{0000-0001-8927-9009}, Julian Holland$^2$ and Rudolf Weber$^{2}$\orcid{0000-0001-8779-2343}}

\affil{$^1$Deutsches Elektronen-Synchrotron DESY, Notkestraße 85, 22607 Hamburg, Germany}

\affil{$^2$Institut für Strahlwerkzeuge, University of Stuttgart, Pfaffenwaldring 43, 70569 Stuttgart, Germany}

\affil{$^*$Author to whom any correspondence should be addressed.}

\email{simon.bohlen@desy.de}




\begin{abstract}
The interaction of ultrafast lasers with plasmas has been studied for many years, primarily with respect to fundamental emission mechanisms. Only in recent years has ionizing radiation emerged as a safety concern in ultrafast laser-based material processing, where high pulse energies, repetition rates, and average powers, combined with continuous material supply, can lead to sustained X-ray emission. These processing-specific findings have informed German radiation protection legislation, which mandates notification or approval for laser systems exceeding irradiances of \SI{1e13}{W/cm^2}. However, this threshold does not distinguish between material processing and other ultrafast laser applications. 
In this work, we show that the conditions required for X-ray generation are highly specific and are typically only met during material processing. We assess the applicability of existing radiation studies to non-processing environments and present experimental results demonstrating negligible or no dose production under representative laboratory conditions, such as ultrafast laser interactions with underdense gas or stationary solid targets. We conclude that current legislation generalizes a processing-specific hazard to all ultrafast laser applications and does not adequately reflect the relevant physical conditions.
\end{abstract}

\section{\label{sec:introduction}Introduction}
Ultrafast lasers, also known as ultra-short pulse lasers, have developed into highly versatile tools, enabling breakthroughs across a wide spectrum of scientific, medical, and industrial fields. Their ability to deliver extremely high peak intensities on ultrashort timescales from picoseconds down to attoseconds has opened entirely new regimes of light–matter interaction. In research, ultrafast lasers underpin ultrafast pump–probe experiments that resolve dynamics of electrons in atoms and molecules \cite{Schultze2010DelayPhotoemission,Isinger2017PhotoionizationDomain}, drive high-harmonic generation \cite{McIntyre1987StudiesGases,Ferray1988Multiple-harmonicGases}, and form the basis of laser–plasma acceleration schemes \cite{Tajima1979LaserAccelerator}. In molecular and condensed matter science, they allow probing and control of processes at the atomic scale and in real time \cite{Dantus1987RealtimeReactions,Zewail1988LaserFemtochemistry}. Ultrafast lasers have also driven new laser technologies such as frequency combs \cite{Udem2002OpticalMetrology} and few-cycle sources \cite{Fork1987CompressionCompensation}, which originated in fundamental research and are rapidly emerging in industry. In laser machining, ultrafast lasers have already revolutionized the industrial processing by enabling surface structuring, medical procedures and advanced manufacturing to unprecedented precision thanks to their highly localized, low-thermal-impact energy deposition \cite{Davis1996WritingLaser,Kawata2001FinerMicrodevices,Malinauskas2016UltrafastIndustry}.

At the same time, the extreme intensities of ultrafast lasers can lead to hot-electron generation and the subsequent emission of X-rays from these energetic electrons during laser–matter interactions \cite{Kuhlke1987SoftPlasmas,Stearns1988GenerationPulses,Murnane1991UltrafastPlasmas,Giulietti1998X-rayPlasmas,Eliezer2002ThePlasmas,Gibbon2005ShortIntroduction,Chen2016IntroductionFusion}. This issue is especially present in the context of high-power industrial material processing, where it was discovered at early stages \cite{Thogersen2001X-rayMicromachining,Bunte2004SecondaryEmission} and gained particular attention with the increase of average powers and repetition rates from modern industrial laser sources \cite{Legall2018X-rayProcessing}. Several studies have shown that, under specific machining conditions, keV-range radiation can be generated, with skin dose rates H$'(0.07)$ reaching values on the order of sieverts per hour in close proximity to the interaction zone inside the machining enclosure \cite{Weber2019ExpectedApplications,Behrens2019X-RAYLASERS,Legall2019TheMachining,Legall2020X-rayProcessing,Freitag2020XRayEnvironment,Legall2021ReviewProcessing,Stolzenberg2021X-raySetting,Schille2021StudyProcessing,Metzner2021X-rayBursts,Mosel2021X-rayDetector,Holland2022InfluenceProcessing,Metzner2026InfluenceAblation}. Importantly, the conditions to produce these dose rates often lie outside the parameter space actually used for productive machining \cite{Weber2022X-rayProblem}.

While these studies convincingly demonstrate that X-ray emission can occur during laser ablation, particularly when high atomic number materials are processed at average powers exceeding tens of watts, the results cover only a narrow range of laser and material parameters and represent a small fraction of the broad field of ultrafast laser applications. Crucially, the exclusive focus on industrial machining scenarios means that the findings cannot simply be transferred to other domains in fundamental research, medical technology, or industrial applications, where the interaction conditions of laser and material are entirely different.

A further limitation is that measured dose rates from machining are often presented only as functions of irradiance, without sufficient discussion of the underlying physics of hot-electron generation and subsequent secondary radiation. As such, German regulatory authorities have introduced rules requiring notification or approval for any laser system capable of delivering irradiances above \SI{1e13}{W/cm^2}, regardless of the specific application or the actual radiation risk. This blanket regulation ignores the diversity of ultrafast laser applications and risks hindering progress in both science and industry by introducing unnecessary bureaucratic barriers.

In this paper, we examine the risk of radiation generation in ultrafast laser applications beyond laser machining. We present experimental data demonstrating that in many applications outside of laser machining no ionizing radiation is measured even at intensities well above \SI{1e13}{W/cm^2}. We also analyze how factors like continuous material replenishment and scanning optics, which are typical in industrial machining but absent in many other ultrafast laser applications, are key drivers of sustained radiation generation.

These observations raise the question of whether a single irradiance-based threshold is sufficient to assess radiation risks across the broad range of ultrafast laser applications. A more differentiated, application-aware evaluation may be necessary to reflect realistic interaction conditions.

In this work, irradiance and intensity are used to denote power per unit area. The term irradiance is primarily employed in the context of laser interaction with material surfaces, whereas intensity refers more generally to peak optical power density, including interactions in gaseous media or cases where no defined surface is present, and therefore irradiance cannot be defined. In addition, both quantities are, where appropriate, also expressed as wavelength-normalized values to account for the quadratic scaling with the laser wavelength, following standard scaling arguments discussed in the literature.


\section{\label{sec:xraygeneration}X-ray generation in laser-plasma interactions}
X-ray emission from laser-produced plasmas has been investigated alongside advances in laser technology, particularly since the invention of chirped-pulse amplification \cite{Strickland1985}. As a result, numerous studies on laser-matter interactions have been published, for example in the context of fusion research, X-ray lithography, and X-ray spectroscopy \cite{Balmer1977ResonancePlasma,Matthews1983CharacterizationRadiography,Corkum1988ThermalExcitation,Weber1988TemporalCavities,Lampart1988ComparativeCalibration,Kim1992Laser-ProducedMicroscopy,Guethlein1996ElectronPulses,Back2003Multi-keVPlasmas,Miaja-Avila2015LaserSpectroscopy}. Therefore, many fundamental aspects are well established and are concisely summarized in many papers and textbooks \cite{Giulietti1998X-rayPlasmas,Gibbon2005ShortIntroduction,Eliezer2002ThePlasmas,Chen2016IntroductionFusion}. At the same time, the interpretation of laser-plasma interactions remains challenging, as X-ray emission depends on a large number of interrelated parameters. Consequently, despite decades of intensive research, a comprehensive and universally applicable description of X-ray generation in laser-plasma interactions is still difficult.

This complexity was concisely summarized by Davies, who stated: "A vast number of papers have been published on laser absorption, containing a wide range of different results and explanations, some of them contradictory, which indicate that absorption depends on almost every laser and target parameter imaginable in a non-trivial manner, making it extremely difficult to give a coherent review of this subject" \cite{Davies2009LaserRegime}. Although this statement was originally made for the specific case of laser absorption in overdense plasmas at relativistic intensities, it captures a more general characteristic of laser-plasma interactions and remains applicable well beyond that regime.

Despite this complexity, the basic mechanisms leading to X-ray emission are well understood. X-ray radiation from laser produced plasmas is closely linked to the generation of hot electrons during the interaction of intense laser pulses with matter. These electrons emit bremsstrahlung radiation upon deceleration in the target or surrounding material and can also induce characteristic X-ray line emission.

Hot electrons can be generated through several absorption mechanisms, including collisional absorption (inverse Bremsstrahlung), resonant absorption, vacuum heating (Brunel absorption), and J×B (ponderomotive) heating \cite{Giulietti1998X-rayPlasmas,Eliezer2002ThePlasmas,Gibbon2005ShortIntroduction}. The relative importance of these mechanisms depends sensitively on plasma density, density gradient, laser incidence angle, polarization, pulse duration, wavelength, intensity, and further interaction parameters.

The plasma density profile plays a particularly important role for efficient hot electron generation. A steep plasma density gradient near the critical surface is often required, where the critical plasma density is given by $n_c = \epsilon_0 m_e \omega_L^2/e^2$, with $\epsilon_0$ denoting the vacuum permittivity, $m_e$ the electron mass, $\omega_L$ the laser frequency, and $e$ the elementary charge. Resonant absorption and vacuum heating mechanisms, in particular, require sharp density gradients at the transition from underdense (below critical density) to overdense (above critical density) plasma regions \cite{Giulietti1998X-rayPlasmas,Eliezer2002ThePlasmas,Gibbon2005ShortIntroduction}. Such conditions are often encountered in laser-based material processing, where solid targets undergo rapid ionization and plasma formation, but are not present in interactions with underdense gas phase plasmas, for example in applications such as spectral broadening or high harmonic generation. 

Efficient energy transfer from the laser to the electrons further depends on laser parameters such as intensity, pulse duration, polarization, wavelength, temporal contrast and more. A universal scaling with intensity alone cannot be identified. Instead, Gibbon summarized the dependence of the hot-electron temperature $T_h$ on the laser intensity $I$ and wavelength $\lambda$ as follows \cite{Gibbon2005ShortIntroduction}: "there is an indisputable general dependence of $T_h$ on the product $I\lambda^2$, rather than on intensity alone, as predicted by all theories". This scaling reflects the dependence on the ponderomotive (quiver) energy ($U_p \propto I \lambda^2$), which is also captured by the normalized vector potential $a_0 = \frac{eE}{m_e c \omega_L}$, where $E$ is the electric field amplitude of the laser and $c$ is the speed of light.

The parameter $a_0$ characterizes relativistic laser intensities when it approaches or exceeds unity, which for wavelengths around \SI{1}{\micro\meter} is achieved at intensities of around \SI{e18}{W/cm^2}. In this relativistic regime, the dominant acceleration mechanisms change, and electrons can be accelerated to MeV energies, either directly by the oscillating electric field of the laser or through the combined action of the laser field and quasi-static electric and magnetic fields generated in the plasma. While the normalized vector potential is commonly used to distinguish between non-relativistic and relativistic interaction regimes, its dependence on both laser intensity and wavelength again illustrates that irradiance alone is insufficient to characterize laser-plasma interactions, thus making it also unsuitable to assess potential radiation hazards.

\subsection{\label{sec:machining}X-ray generation in industrial laser machining}
Industrial ultrafast laser material processing is typically performed at irradiances below \SI{e13}{W/cm^2} in order to optimize processing quality and efficiency \cite{Neuenschwander2010,Lauer2014,Weber2022X-rayProblem}. Nevertheless, the possibility of X-ray generation at higher irradiances, exceeding \SI{e13}{W/cm^2}, has been the focus of several studies in recent years \cite{Weber2019ExpectedApplications,Behrens2019X-RAYLASERS,Legall2019TheMachining,Legall2020X-rayProcessing,Freitag2020XRayEnvironment,Legall2021ReviewProcessing,Stolzenberg2021X-raySetting,Schille2021StudyProcessing,Metzner2021X-rayBursts,Mosel2021X-rayDetector,Holland2022InfluenceProcessing,Metzner2026InfluenceAblation}.
In these works, X-ray emission is associated with machining conditions involving high pulse energies (100 µJ–mJ), high repetition rates (hundreds of kHz to MHz), and dense, high-Z materials such as tungsten, copper, or steel. The combination of high pulse energy and repetition rate results in substantial average powers, enabling sustained energy deposition and the formation of persistent plasmas that promote hot-electron generation. However, radiation yield and dose are frequently discussed mainly in terms of irradiance or pulse energy, without explicit consideration of the established scaling of hot-electron heating with the ponderomotive quiver energy and the parameter $I\lambda^2$, although notable exceptions exist \cite{Weber2019ExpectedApplications}. This risks obscuring physical mechanisms of X-ray generation that are well understood from laser–plasma interaction research in high-energy-density physics, inertial confinement fusion, and laser-driven radiation source development.

This disconnect becomes particularly relevant when apparently new effects are reported, for example increased dose rates under burst-mode operation. Interpretation within the framework of established laser–plasma theory, such as considering plasma gradient formation, residual plasma between sub-pulses, and the resulting changes in absorption and coupling, could provide physically grounded explanations of such observations. Without this context, experimental findings risk being framed primarily in terms of unexpectedly high or potentially alarming dose levels, rather than as consequences of modified interaction conditions. At the same time, theory-guided interpretation would help to identify which aspects of plasma formation and energy coupling warrant targeted investigation using additional diagnostics to broaden the understanding of the changed coupling effects.

Admittedly, the interpretation of many material-processing experiments is challenging, as these experiments operate in an intermediate interaction regime in which different absorption mechanisms may dominate, depending not only on irradiance but also on angle of incidence, polarization, pulse duration, temporal contrast, pre-plasma scale length, evolving surface and plasma conditions. However, in such transitional regimes, interpretation within appropriate physical models is essential to enable meaningful scaling of the results to other experimental conditions. Established and calibrated models to estimate radiation emission in laser machining scenarios already exist in the literature \cite{Weber2019ExpectedApplications,Holland2022InfluenceProcessing}.

Finally, some experimental and modeling studies explore parameter regimes that extend well beyond those compatible with efficient or practical ultrafast laser-based material processing \cite{Weber2022X-rayProblem}, in particular when approaching irradiances of \SI{e16}{\watt\per\centi\metre\squared \micro\meter\squared}. While such studies are valuable for investigating fundamental emission mechanisms and upper-bound conditions, their direct transfer to industrial machining environments is limited, especially when dose rates are evaluated at positions, that are usually in strongly confined or fully enclosed geometries. A clearer distinction between realistic processing conditions and extreme parameter studies would therefore improve both physical interpretation and the relevance of radiation safety assessments for industrial applications.

An important factor that distinguishes material processing from many other applications of ultrafast lasers is continuous material renewal at the interaction site. While this aspect was already recognized at an early stage \cite{Thogersen2001X-rayMicromachining}, its role in enabling conditions under which sustained and elevated dose rates can occur, by maintaining a persistent source of dense plasma and energetic electrons, appears to receive less attention in more recent discussions. It is therefore important to emphasize that consistently high dose rates are not determined by individual laser parameters alone, but rely on the combined effect of laser settings and dynamic target replenishment inherent to machining processes.

\subsection{\label{sec:plasma_science}Potential X-ray generation in other ultrafast laser applications}
Beyond industrial laser machining, ultrafast lasers are increasingly employed in scientific and technological applications operating at peak intensities that often reach or exceed \SI{e13}{W/cm^2}, while presenting a negligible risk of X-ray emission (> \SI{5}{keV}). In these regimes, the laser–matter interaction avoids the formation of sustained hot plasmas through the use of gaseous media, sub-critical density targets, or shorter wavelengths (e.g. UV), which suppress hot-electron generation due to reduced ponderomotive scaling, as discussed above.

Prominent examples include nonlinear spectral broadening and post-compression techniques, which are now transitioning from laboratory environments to industrial manufacturing. These processes employ noble gases in multipass (Herriott) cells \cite{Rajhans2023Post-compressionRegime} or hollow-core fibers \cite{Nagy2019GenerationPower} to broaden the laser spectrum via self-phase modulation, inherently precluding efficient electron heating and the generation of high-energy radiation.

Similarly, high-harmonic generation (HHG) utilizes intensities exceeding \SI{e13}{W/cm^2} in dilute gases to produce coherent VUV and EUV radiation \cite{McIntyre1987StudiesGases,Ferray1988Multiple-harmonicGases}, while remaining intrinsically free of X-ray emission. In research, this enables attosecond science \cite{Krausz2009AttosecondPhysics} and angle-resolved photoemission spectroscopy for mapping electronic band structures in quantum materials \cite{Zhang2022Angle-resolvedSpectroscopy}. Industrially, HHG-based sources have been adopted for EUV lithography mask inspection and metrology, where short wavelengths enable the detection of microscopic defects on silicon wafers \cite{Gardner2014TabletopPtychography,Gardner2017SubwavelengthSource}. In addition, HHG provides a “soft” ionization source for photo-ionization mass spectrometry, allowing fragment-free analysis of complex organic molecules in environmental and chemical monitoring applications \cite{Worner2011ConicalSpectroscopy}.

Femtosecond UV laser-induced breakdown spectroscopy operated at intensities on the order of \SI{e13}{W/cm^2} has also been demonstrated as a diagnostic tool for real-time, depth-resolved monitoring of micro-LED structures, where the high peak intensity of femtosecond pulses improves signal-to-noise ratio and analytical precision \cite{Jung2025InterfaceSpectroscopy}.

Ultrashort-pulse lasers may further interact with materials either within the laser chain itself or at external targets. For example, pulse-cleaning techniques can involve interactions with pinhole targets even at very high intensities \cite{Murray1997SpatialIssues}. Similarly, high-intensity lasers have been used to drill apertures into thin windows for differential pumping, thereby ensuring optimal alignment between the laser beam and the aperture \cite{Steingrube2009PhaseCell,Appi2021SynchronizedStudies}. In such configurations, the absence of continuous material renewal limits cumulative plasma heating and thus mitigates sustained X-ray production.

Finally, femtosecond pulses with peak intensities exceeding \SI{e13}{W/cm^2} are also employed for high-field terahertz generation. In particular, two-color laser-induced plasma filaments reach very high instantaneous intensities. However, the low-density nature of the gaseous medium restricts the formation of high-energy electron populations \cite{Hochstrasser2000IntenseAir}. Terahertz diagnostics play an increasingly important role for future key technologies, such as semiconductor manufacturing and battery production, by enabling non-destructive, depth-resolved quality control of increasingly complex material stacks and interfaces \cite{Jepsen2011TerahertzApplications}.

The examples discussed above are representative of a broad and rapidly expanding class of ultrafast laser applications operating at very high peak intensities without significant X-ray emission. Ongoing advances in laser technology, nonlinear optics, and diagnostic methodologies continue to extend the range of use cases for ultrafast lasers in both research and industrial environments, reinforcing their role as versatile high-field tools that can be deployed safely when appropriate interaction regimes are selected.

\section{\label{sec:experimental_setup}Measurements of X-ray generation beyond laser machining}
\begin{figure}[htb]
\centering
\includegraphics[width=0.95\columnwidth]{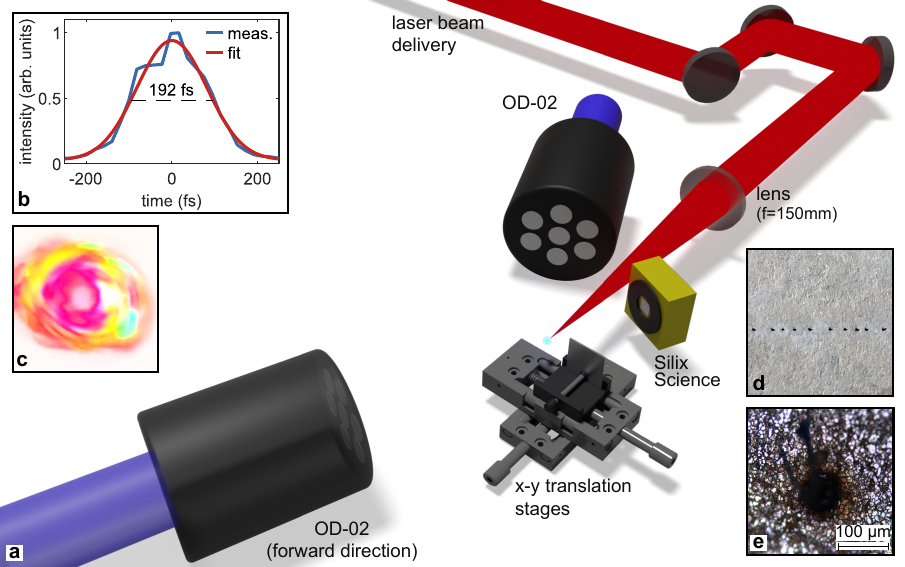}
\caption{
    Experimental setup and representative diagnostic images.
    \textbf{a}, Schematic overview of the experimental setup for dose estimation. The laser beam is focused by a lens to generate either an air plasma or to drill holes in stationary overcritical targets at normal incidence. The targets are mounted on translation stages for alignment. X-rays are measured using an OD-02 detector and a Silix Science detector positioned at distances of \SI{20}{cm} and \SI{10}{cm}, respectively. For overcritical targets (tungsten and steel), a second OD-02 detector was placed in the forward direction (in laser propagation direction).
    \textbf{b}, Pulse duration measurement obtained with an autocorrelator.
    \textbf{c}, Image of plasma emission projected onto a screen placed behind the air plasma.
    \textbf{d}, Optical image of ten holes drilled into a \SI{1}{mm} thick tungsten target.
    \textbf{e}, Microscope image of a hole drilled into a steel target.
}
\label{fig:expsetup}
\end{figure}

To estimate potential radiation doses arising from ultrafast laser applications beyond industrial laser machining, a series of dedicated experiments was performed at the Institut für Strahlwerkzeuge (IFSW) of the University of Stuttgart. The experimental configurations were selected to represent typical high-intensity use-case scenarios, while deliberately excluding continuous material feed (see Fig.~\ref{fig:expsetup}d).

Two main considerations motivated the present study. First, the role of underdense plasmas, which are routinely employed in many ultrafast laser applications, should be further investigated even at high peak intensities. Second, the influence of material renewal on the generation of ionizing radiation in ultrafast laser interactions requires clarification. Previous studies have reported that little or no radiation is emitted from stationary targets \cite{Thogersen2001X-rayMicromachining}, as the material located at the high-intensity beam waist is rapidly ablated, thereby suppressing sustained plasma formation and causing the effective irradiance at the interaction region to decrease as the target surface recedes.

The experiments presented here were therefore designed to systematically investigate stationary interaction scenarios and to establish quantitative upper bounds on the resulting radiation doses. In doing so, they refine earlier conclusions by demonstrating not only the absence of sustained radiation generation, but also by providing absolute dose estimates for well-defined worst-case configurations.

The experimental setup used for these measurements is shown in Fig.~\ref{fig:expsetup}.

The laser source was a Ti:Sapphire chirped-pulse amplification (CPA) system operating at a central wavelength of \SI{800}{nm} and a repetition rate of \SI{1}{kHz}. The average output power was \SI{3.8}{W}, corresponding to a pulse energy of \SI{3.8}{mJ}. The pulse duration was measured using an autocorrelator to be \SI{192 \pm 2}{fs} (FWHM, see Fig.~\ref{fig:expsetup}d).

Prior to focusing, the beam diameter was \SI{10}{mm} (1/e$^2$). Focusing was achieved using a lens with a focal length of \SI{150}{mm}, resulting in a calculated minimum beam radius of approximately \SI{8.4}{\um} (assuming $M^2 = 1.1$). Under ideal conditions, this yields a nominal peak intensity of up to \SI{1.7e16}{W/cm^2} or a wavelength-normalized intensity $I\lambda^2$ = \SI{1.1e16}{\watt\per\centi\metre\squared \micro\meter\squared}. However, the measured temporal profile exhibited noticeable shoulders, most likely caused by residual fourth-order dispersion (Fig.~\ref{fig:expsetup}b). In addition, ionization-induced defocusing in air is expected to noticeably reduce the effective peak intensity at focus \cite{Gibbon2005ShortIntroduction,Weber2022X-rayProblem}, consistent with the plasma emission pattern observed in Fig.~\ref{fig:expsetup}c. The true peak intensity (irradiance) is therefore expected to be lower than this upper-bound estimate. Exact in-situ intensity measurements are challenging at the intensities considered here, and consequently the peak intensity at the interaction point cannot be determined with absolute precision. This uncertainty is generally implicit in comparable industrial laser machining studies but is stated explicitly here to clearly delineate the conservative bounds of the present analysis.

Ionizing radiation was monitored using two complementary detector systems. An OD-02 detector was positioned at a distance of \SI{20}{cm} from the interaction point at an angle of \SI{155}{\degree} relative to the laser propagation direction. In addition, a Silix Science detector was placed at a distance of \SI{10}{cm} at an angle of \SI{205}{\degree}. Both detectors are suitable for measurements of the directional dose equivalent H$'(0.07)$ and have been employed extensively in previous ultrafast laser experiments, where they are commonly used to monitor ultra-short X-ray pulses generated in laser–matter interactions. The detector positions were selected to probe near-backscattering geometries, where the highest dose rates are typically expected, while avoiding obstruction of the incident laser beam and minimising shielding effects caused by target geometry or material redeposition known from hole-drilling \cite{Holland2024Self-ShieldingZone}.

For experiments involving overcritical densities (i.e., non-transparent solid targets), an additional OD-02 detector was positioned in the forward direction at an angle of \SI{20}{\degree} and a distance of \SI{20}{cm} from the interaction point. This forward-scattering geometry provided complementary information on the angular distribution of emitted radiation. A transverse stage allowed lateral repositioning of the target between individual measurements without altering the nominal focal conditions, while a second stage along the laser propagation direction enabled alignment of the target with the focal plane (Fig.~\ref{fig:expsetup}a).

\subsection{\label{sec:air_exps}X-ray measurements in air}
To investigate potential X-ray emission from gas-phase laser--matter interactions, the laser was focused into ambient air with nominal wavelength-normalized intensities $I\lambda^2$ of up to \SI{1.1e16}{\watt\per\centi\metre\squared \micro\meter\squared}. Measurements were conducted in multiple sets with durations of several minutes to minimize sensitivity to short-term fluctuations and long-term drifts of the laser system. Background levels were determined with the laser switched off while maintaining identical detector positions and geometry.

\begin{figure}[H]
\centering
\includegraphics[width=0.95\columnwidth]{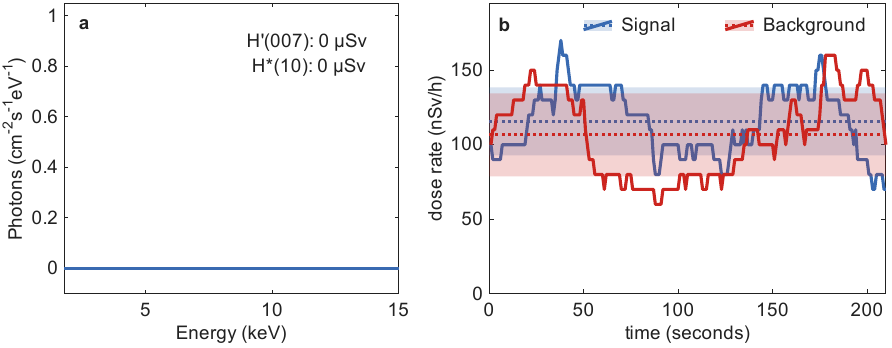}
\caption{X-ray measurements in air at nominal wavelength-normalized intensities $I\lambda^2$ of up to \SI{e16}{\watt\per\centi\metre\squared \micro\meter\squared}. 
    \textbf{a}, Photon spectrum and absolute dose measured with the Silix Science detector during laser operation. 
    \textbf{b}, Dose rate measured with the OD-02 detector with laser operation (signal) and without laser operation (background).}
\label{fig:specair}
\end{figure}

The photon spectrum and absolute dose measured with the Silix Science detector are shown in Fig.~\ref{fig:specair}a. No photon counts above the detector background threshold were registered in any of the measurements. Given the close proximity of the detector to the interaction region, this provides strong evidence that no measurable X-rays with photon energies above the detector threshold (>~\SI{2}{keV}) were generated under these conditions. In particular, X-ray emission above \SI{5}{keV} can be excluded.

The corresponding dose rate measured with the OD-02 detector is shown in Fig.~\ref{fig:specair}b. The measured dose rate is indistinguishable from the detector background and shows no statistically significant deviation when the laser is operated at its full power. This behaviour was consistently observed across all measurement runs.

These results demonstrate that interactions of high-intensity laser pulses with underdense plasmas (in this case ambient air) at nominal wavelength-normalized peak intensities up to approximately \SI{e16}{\watt\per\centi\metre\squared \micro\meter\squared} do not lead to detectable X-ray emission and remain well below the detection limits of standard radiation monitoring equipment, in agreement with theoretical expectations.

At substantially higher intensities, however, the interaction dynamics change fundamentally. As relativistic laser intensities are approached ($a_0 \geq 1$), new coupling mechanisms become accessible and the generation of high-energy electrons in underdense plasmas becomes possible. These regimes form the basis of laser–plasma accelerators \cite{Tajima1979LaserAccelerator}, where electron bunches with energies in the 10 GeV range can be produced \cite{Picksley2024MatchedAccelerators}. Similarly, when femtosecond laser pulses are tightly focused in air at relativistic intensities, MeV-scale electron energies and correspondingly high radiation dose rates have been reported \cite{Vallieres2024HighAir}.

\subsection{\label{sec:stationary_exps}X-ray measurements for stationary targets}
To investigate potential absolute doses generated by laser interactions with overcritical densities (i.e., absorbing solid-density materials) in the absence of material feed, experiments were performed on stationary tungsten and steel targets. These materials were chosen, as they represent conservative worst-case scenarios in typical machining investigations \cite{Legall2021ReviewProcessing}. 

\begin{figure}[H]
\centering
\includegraphics[width=0.90\columnwidth]{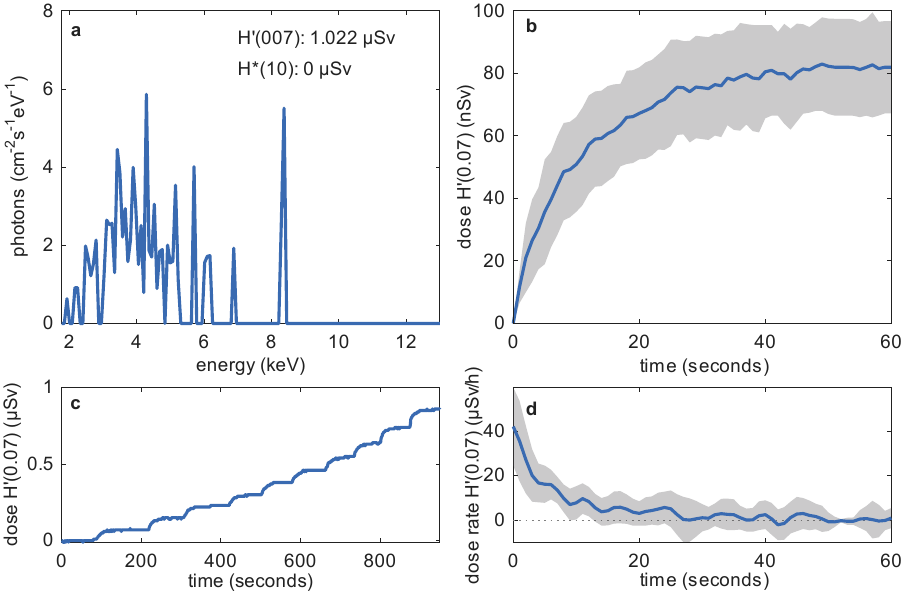}
\caption{Spectrum and absolute doses measured during drilling of \SI{1}{mm} thick, stationary tungsten targets at nominal wavelength-normalized laser irradiance $I\lambda^2$ of up to \SI{e16}{\watt\per\centi\metre\squared \micro\meter\squared}. 
    \textbf{a}, Spectrum and absolute dose integrated for all ten holes measured with the Silix Science detector.
    \textbf{b}, Evolution of the average dose per hole with standard deviation measured with the OD-02 detector.
    \textbf{c}, Evolution of the total integrated dose for all ten holes measured with the OD-02 detector.
    \textbf{d}, Temporal evolution of the average dose rate with standard deviation, obtained by numerical differentiation of the data shown in panel b. Mild smoothing (three-point moving average) was applied to mitigate noise amplification inherent to numerical differentiation. The dotted line indicates the zero line.
}
\label{fig:doses_tungsten}
\end{figure}

For tungsten, a \SI{1}{mm} thick plate was placed at the laser focus and aligned using a longitudinal translation stage. Initial alignment was performed at reduced laser intensities to avoid premature ablation. It is noted, however, that ionization-induced defocusing may lead to shifts of the effective focus position during high-intensity operation as discussed above. Ten holes were drilled at the maximum nominal wavelength-normalized laser irradiance $I\lambda^2$ = \SI{1.1e16}{\watt\per\centi\metre\squared \micro\meter\squared}, with both the laser and the target kept stationary for each hole (aside from unavoidable beam jitter). The results are summarized in Fig.~\ref{fig:doses_tungsten}.

Figure~\ref{fig:doses_tungsten}a shows the integrated X-ray spectrum and total dose recorded with the Silix Science detector for all ten holes. The mean dose per hole was \SI{0.10 \pm 0.05}{\micro Sv} (H$'(0.07)$), with individual values ranging from \SI{0.03}{\micro Sv} to \SI{0.2}{\micro Sv}. Owing to the low cumulative dose of approximately \SI{1}{\micro\sievert}, the resulting spectrum exhibits considerable statistical noise. Despite this, the spectrum resembles a thermal distribution, with the low-energy part attenuated by air between the plasma and the detector, resulting in a peak around \SI{4}{keV}. A distinctive line feature around \SI{8}{keV} is most likely attributed to $L\alpha$ emission, expected in the range of \SIrange{8.3}{8.4}{keV} for tungsten.

The corresponding OD-02 measurements are presented in Fig.~\ref{fig:doses_tungsten}b--d. The average dose per hole was \SI{82 \pm 15}{nSv} (H$'(0.07)$), slightly below the Silix value, likely due to differences in detector distance and thus increased air attenuation. The cumulative dose for all ten holes (Fig.~\ref{fig:doses_tungsten}c), recorded in a single continuous measurement, amounted to approximately \SI{0.85}{\micro\sievert}.

Numerical differentiation of the individual OD-02 traces yields the averaged temporal dose rate shown in Fig.~\ref{fig:doses_tungsten}d. Mild smoothing (three-point moving average) was applied to this panel to mitigate noise amplification inherent to differentiation. The initial dose rate reaches \SI{42 \pm 17}{\micro\sievert\per\hour} and decreases rapidly, oscillating around the zero line after about \SI{27}{s}. Although each hole was drilled through the \SI{1}{mm} plate within approximately \SI{60}{s}, the pronounced reduction in dose rate after roughly the first half of the drilling time is attributed to increasing self-shielding by the surrounding material and a significant decrease in the effective irradiance within the developing channel.

The absolute doses measured for tungsten are not expected to increase substantially at higher repetition rates or for thicker targets. While higher repetition rates reduce the drilling time, the number of laser pulses required to produce a single hole remains approximately constant, and thus the total X-ray yield per hole is not expected to change significantly. For thicker materials, increased self-shielding and deviations from optimal focusing conditions are expected to further limit the detectable radiation. The presented results therefore represent conservative worst-case scenarios, including misalignment or unintended stationary exposure of solid material in a high-intensity laser focus. Even under these conditions, the detected radiation levels remain in the nanosievert range and thus negligible, indicating that the associated ionizing radiation hazard is minimal.

\begin{figure}[H]
\centering
\includegraphics[width=0.90\columnwidth]{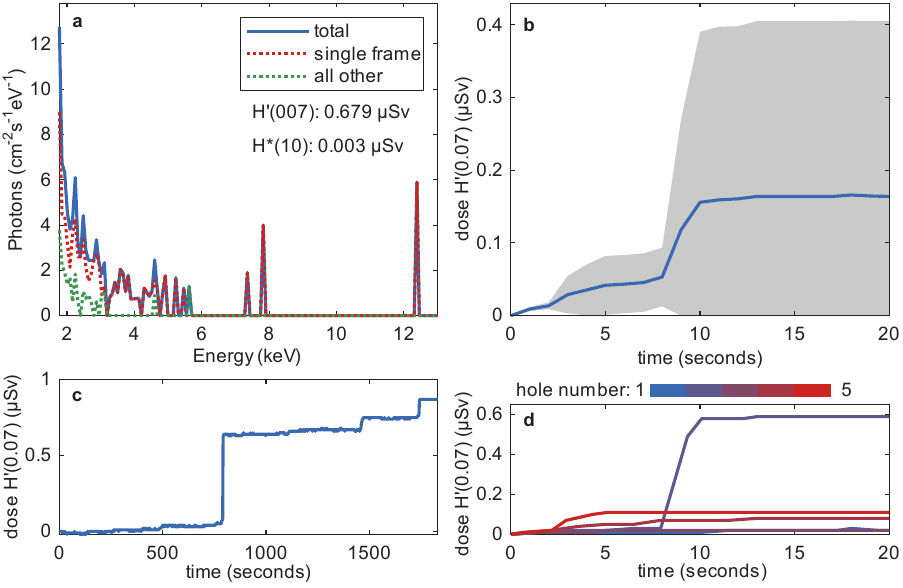}
\caption{Spectrum and absolute doses measured during drilling of \SI{3}{mm} thick, stationary steel targets at nominal laser pulse intensities of up to \SI{e16}{\watt\per\centi\metre\squared \micro\meter\squared}. 
    \textbf{a}, Spectrum and integrated absolute dose for all five holes measured with the Silix Science detector.
    \textbf{b}, Evolution of the average dose per hole with standard deviation measured with the OD-02 detector.
    \textbf{c}, Evolution of the total integrated dose for all five holes measured with the OD-02 detector.
    \textbf{d}, Evolution of the absolute dose for each of the five individual holes measured with the OD-02 detector.
}
\label{fig:doses_steel}
\end{figure}

A second set of measurements was performed using \SI{3}{mm}-thick steel targets at the same nominal wavelength-normalized laser irradiance $I\lambda^2$ = \SI{1.1e16}{\watt\per\centi\metre\squared\micro\metre\squared}, where five stationary holes were drilled. In contrast to the reproducible behavior observed for tungsten, the measured doses exhibit pronounced shot-to-shot variations despite nominally identical experimental conditions. The dose per hole varies by more than an order of magnitude, ranging from approximately \SI{20}{nSv} to \SI{600}{nSv}. This pronounced variability demonstrates that the X-ray emission cannot be described by a single interaction parameter alone, but is governed by highly non-linear and dynamically evolving processes. Despite this, all observed dose values remain radiologically insignificant. The results are summarized in Fig.~\ref{fig:doses_steel}.

Fig.~\ref{fig:doses_steel}a shows the integrated X-ray spectrum and total absolute dose measured with the Silix Science detector for all five holes. The majority of detected photons originate from a single measurement interval of approximately \SI{6}{s} and exhibit a strong contribution of photons with energies below \SI{3}{keV}. This is unexpected, as this part of the spectrum should be strongly attenuated by air between the plasma and the detector, similar to the case of tungsten. The remaining spectral contribution was accumulated over the total drilling time of approximately \SI{100}{s} required to produce all five holes. Although the overall acquisition time extended over several minutes, photon counts were recorded exclusively during active drilling (approximately \SI{20}{s} per hole). The data therefore indicate that the dominant emission occurred during a short transient event.

The corresponding OD-02 data are shown in Fig.~\ref{fig:doses_steel}b--d. The average dose per hole (Fig.~\ref{fig:doses_steel}b) exhibits stronger fluctuations than observed for tungsten, consistent with the larger variance in the individual hole evolution shown in Fig.~\ref{fig:doses_steel}d. Two holes yielded doses of approximately \SI{20}{nSv}, two around \SI{100}{nSv}, and one significantly higher value of approximately \SI{600}{nSv}. This pronounced transient event is clearly visible both in the cumulative dose curve (Fig.~\ref{fig:doses_steel}c) and in the single-hole evolution (Fig.~\ref{fig:doses_steel}d), and corresponds to the dominant Silix measurement interval discussed above.

The integrated dose of \SI{680}{\nano\sievert} measured with the Silix detector is slightly lower than the \SI{870}{\nano\sievert} derived from the cumulative OD-02 curve shown in Fig.~\ref{fig:doses_steel}c. Considering the larger detector distance of the OD-02 and the predominantly low photon energies observed in the spectrum, a lower OD-02 signal would in principle be expected. Within the combined statistical and systematic uncertainties of these measurements, the deviation is therefore attributed to limited photon statistics.

The temporal evolution of the dose for individual holes shown in Fig.~\ref{fig:doses_steel}d demonstrates that the enhanced emission associated with the pronounced transient event did not occur at the onset of drilling, but only after approximately \SI{8}{s}. A similar, though less pronounced, delayed increase is observed for a second hole, appearing approximately \SI{2}{s} after drilling commenced.

The origin of these transient and strongly non-linear enhancements cannot be unambiguously identified based on the available data. Several mechanisms may contribute, including shot-to-shot variations in beam pointing or target positioning, which may locally increase the effective irradiance within the developing channel. Mechanical vibrations of the setup may further modify the relative position between focus and target, affecting material replenishment and plasma formation. In addition, dynamic changes in laser–plasma coupling within the borehole, including enhanced absorption in confined geometries \cite{Gibbon2005ShortIntroduction}, may transiently increase hot-electron generation and bremsstrahlung yield. Evolving plasma density gradients and interaction geometry may also influence the emission efficiency over time.

Given the limited statistics, the relative contribution of these mechanisms cannot be resolved. Nevertheless, even the most pronounced transient events remain at radiologically negligible levels. Despite the higher average doses observed for steel compared to tungsten, all measured values remain in the nanosievert range. Even the maximum recorded value of \SI{600}{nSv} (H$'(0.07)$) is well below typical daily natural background levels, which are on the order of a few \SI{}{\micro Sv} in terms of ambient dose equivalent H$^*(10)$. From a radiation protection perspective, these doses are insignificant and far below any level of concern. The results therefore confirm that stationary irradiation of common engineering materials with high-intensity laser pulses does not pose a relevant radiological hazard.

The additional OD-02 detector positioned in forward direction did not record dose values significantly above background for any of the solid-target measurements. This is consistent with strong self-shielding of the material in forward direction, where the bulk of the target attenuates bremsstrahlung generated within the interaction region. The absence of measurable forward emission further indicates that the detector geometries employed in this study provide conservative upper-bound estimates for dose assessment.

Nevertheless, the experiments presented here are subject to several inherent limitations. Most notably, the exact peak irradiance at the interaction point cannot be determined with high precision due to ionization-induced defocusing and dynamic changes of the focal conditions during laser operation. While nominal intensities were estimated based on beam and pulse parameters, the true peak irradiance remains uncertain. In addition, for solid targets, alignment to the laser focus was performed at reduced intensities, and shifts of the effective focus position during high-intensity irradiation cannot be excluded.

Despite these limitations, the experimental results allow for robust qualitative and quantitative conclusions. In particular, the measurements demonstrate that stationary targets exposed to high-intensity ultra-short laser pulses produce only transient and strongly self-limited radiation emission. While the tungsten results indicate a largely self-limiting and reproducible emission behavior, the steel measurements demonstrate that the underlying interaction process can exhibit strong non-linear and stochastic characteristics under nominally identical conditions. The absolute doses measured for tungsten and steel, ranging from several tens of nanosieverts up to a few hundred nanosieverts per drilled hole, confirm and quantify earlier observations that rapid ablation of material at the focal spot effectively suppresses sustained X-ray generation \cite{Thogersen2001X-rayMicromachining}.

These findings underline that continuous material renewal, such as that occurring in laser machining with moving targets, material feed, or persistent plasma interaction volumes, is a key driver for sustained radiation generation in ultrafast laser applications. By placing numerical dose values on stationary interaction scenarios, the present study refines earlier qualitative statements and provides a quantitative basis for radiation safety assessments in research environments.

\section{\label{sec:Summary}Summary}
This paper investigates the generation of X-rays in ultrafast laser applications and critically examines the use of a fixed irradiance threshold of \SI{e13}{W/cm^2} in radiation protection regulations. While previous studies have shown that hazardous X-ray emission can occur during industrial ultrafast laser-based material processing, these findings are largely restricted to machining scenarios involving high average powers, pulse energies, and repetition rates, as well as high-Z materials and continuous material renewal.

Using established laser–plasma interaction theory, we show that X-ray emission is governed by hot-electron generation and plasma-coupling conditions, depending on many process parameters, rather than irradiance alone. While the inclusion of parameters such as the wavelength (e.g. through $I\lambda^2$) allows general trends to be captured, even this parameterization is insufficient to assess radiation hazards, as the comparison of under- and overcritical densities demonstrates that the interaction cannot be reduced to a small set of parameters. Nevertheless, sustained radiation production at the intensities of interest requires persistent and overdense plasmas and is therefore generally not expected outside material-processing environments.

To assess radiation risks applicable to many ultrafast laser application in industrial or research settings, we performed dedicated measurements using a high-intensity femtosecond Ti:Sapphire laser. No ionizing radiation above background was detected for laser interaction with ambient air at nominal wavelength-normalized laser intensity $I\lambda^2$ of up to 
\SI{e16}{\watt\per\centi\metre\squared \micro\meter\squared}. For stationary tungsten and steel targets, only transient radiation was observed, with absolute directional doses H$'(0.07)$ in the nanosievert range per interaction, thus far below radiological relevance.

Both experimental results and theoretical considerations demonstrate that irradiance alone is not a suitable metric for assessing radiation risk in ultrafast laser applications. Radiation protection regulations should therefore be based on realistic radiological risk under specific interaction conditions rather than on a single laser parameter. An application-aware framework would ensure proportional radiation protection while maintaining safe working conditions and supporting continued scientific and technological development.

\section*{\label{sec:Acknowledgements}Acknowledgements}
The authors used generative AI (ChatGPT and Gemini) in the preparation of this manuscript for language editing and drafting. Following the use of this tool, the authors reviewed and edited the content as needed and take full responsibility for the final content of the manuscript.

\bibliographystyle{apsrev4-2}
\bibliography{references}

@article{Strickland1985,
    title = {{Compression of Ampified Chirped Optical Pulses}},
    year = {1985},
    journal = {Optics Communications},
    author = {Strickland, D. and Mourou, G.},
    number = {6},
    pages = {447--449},
    volume = {55},
    isbn = {0030-4018},
    doi = {10.1016/0030-4018(85)90151-8},
    issn = {00304018}
}

@article{Zhang2022Angle-resolvedSpectroscopy,
    title = {{Angle-resolved photoemission spectroscopy}},
    year = {2022},
    journal = {Nature Reviews Methods Primers 2022 2:1},
    author = {Zhang, Hongyun and Pincelli, Tommaso and Jozwiak, Chris and Kondo, Takeshi and Ernstorfer, Ralph and Sato, Takafumi and Zhou, Shuyun},
    number = {1},
    month = {7},
    pages = {54-},
    volume = {2},
    publisher = {Nature Publishing Group},
    url = {https://www.nature.com/articles/s43586-022-00133-7},
    isbn = {0123456789},
    doi = {10.1038/s43586-022-00133-7},
    issn = {2662-8449},
    arxivId = {2207.06942}
}

@article{Krausz2009AttosecondPhysics,
    title = {{Attosecond physics}},
    year = {2009},
    journal = {Reviews of Modern Physics},
    author = {Krausz, Ferenc and Ivanov, Misha},
    number = {1},
    month = {1},
    pages = {163--234},
    volume = {81},
    publisher = {American Physical Society},
    url = {https://journals.aps.org/rmp/abstract/10.1103/RevModPhys.81.163},
    doi = {10.1103/REVMODPHYS.81.163/FIGURES/36/MEDIUM},
    issn = {15390756}
}

@article{Matthews1983CharacterizationRadiography,
    title = {{Characterization of laser-produced plasma x-ray sources for use in x-ray radiography}},
    year = {1983},
    journal = {Journal of Applied Physics},
    author = {Matthews, D. L. and Campbell, E. M. and Ceglio, N. M. and Hermes, G. and Kauffman, R. and Koppel, L. and Lee, R. and Manes, K. and Rupert, V. and Slivinsky, V. W. and Turner, R. and Ze, F.},
    number = {8},
    month = {8},
    pages = {4260--4268},
    volume = {54},
    publisher = {AIP Publishing},
    doi = {10.1063/1.332680},
    issn = {00218979}
}

@misc{Lampart1988ComparativeCalibration,
    title = {{Comparative x-ray spectroscopy of various-Z elements with wavelength calibration}},
    year = {1988},
    booktitle = {Journal of Applied Physics},
    author = {Lampart, W. and Weber, R. and Balmer, J. E.},
    number = {2},
    month = {1},
    pages = {273--276},
    volume = {63},
    publisher = {AIP Publishing},
    doi = {10.1063/1.340287},
    issn = {00218979}
}

@article{Fork1987CompressionCompensation,
    title = {{Compression of optical pulses to six femtoseconds by using cubic phase compensation}},
    year = {1987},
    journal = {Optics Letters, Vol. 12, Issue 7, pp. 483-485},
    author = {Fork, R. L. and Becker, P. C. and Cruz, C. H. Brito and Shank, C. V.},
    number = {7},
    month = {7},
    pages = {483--485},
    volume = {12},
    publisher = {Optica Publishing Group},
    url = {https://opg.optica.org/viewmedia.cfm?uri=ol-12-7-483&seq=0&html=true https://opg.optica.org/abstract.cfm?uri=ol-12-7-483 https://opg.optica.org/ol/abstract.cfm?uri=ol-12-7-483},
    doi = {10.1364/OL.12.000483},
    issn = {1539-4794},
    pmid = {19741772}
}

@article{Worner2011ConicalSpectroscopy,
    title = {{Conical intersection dynamics in NO2 probed by homodyne high-harmonic spectroscopy}},
    year = {2011},
    journal = {Science},
    author = {W{\"{o}}rner, H. J. and Bertrand, J. B. and Fabre, B. and Higuet, J. and Ruf, H. and Dubrouil, A. and Patchkovskii, S. and Spanner, M. and Mairesse, Y. and Blanchet, V. and M{\'{e}}vel, E. and Constant, E. and Corkum, P. B. and Villeneuve, D. M.},
    number = {6053},
    month = {10},
    pages = {208--212},
    volume = {334},
    publisher = {American Association for the Advancement of Science},
    url = {/doi/pdf/10.1126/science.1208664?download=true},
    doi = {10.1126/SCIENCE.1208664;PAGE:STRING:ARTICLE/CHAPTER},
    issn = {10959203}
}

@article{Schultze2010DelayPhotoemission,
    title = {{Delay in photoemission}},
    year = {2010},
    journal = {Science},
    author = {Schultze, M. and Fie{\ss}, M. and Karpowicz, N. and Gagnon, J. and Korbman, M. and Hofstetter, M. and Neppl, S. and Cavalieri, A. L. and Komninos, Y. and Mercouris, Th and Nicolaides, C. A. and Pazourek, R. and Nagele, S. and Feist, J. and Burgd{\"{o}}rfer, J. and Azzeer, A. M. and Ernstorfer, R. and Kienberger, R. and Kleineberg, U. and Goulielmakis, E. and Krausz, F. and Yakovlev, V. S.},
    number = {5986},
    month = {6},
    pages = {1658--1662},
    volume = {328},
    publisher = {American Association for the Advancement of Science},
    url = {/doi/pdf/10.1126/science.1189401},
    doi = {10.1126/SCIENCE.1189401;WEBSITE:WEBSITE:AAAS-SITE;JOURNAL:JOURNAL:SCIENCE;WGROUP:STRING:PUBLICATION},
    issn = {00368075}
}

@article{Guethlein1996ElectronPulses,
    title = {{Electron temperature measurements of solid density plasmas produced by intense ultrashort laser pulses}},
    year = {1996},
    journal = {Physical Review Letters},
    author = {Guethlein, G. and Foord, M. E. and Price, D.},
    number = {6},
    month = {8},
    pages = {1055--1058},
    volume = {77},
    publisher = {American Physical Society},
    doi = {10.1103/PhysRevLett.77.1055},
    issn = {10797114}
}

@article{Weber2019ExpectedApplications,
    title = {{Expected X-ray dose rates resulting from industrial ultrafast laser applications}},
    year = {2019},
    journal = {Applied Physics A: Materials Science and Processing},
    author = {Weber, Rudolf and Giedl-Wagner, Roswitha and F{\"{o}}rster, Daniel J. and Pauli, Anton and Graf, Thomas and Balmer, Jürg E.},
    number = {9},
    month = {9},
    pages = {1--12},
    volume = {125},
    publisher = {Springer Verlag},
    doi = {10.1007/s00339-019-2885-1},
    issn = {14320630}
}

@article{Kawata2001FinerMicrodevices,
    title = {{Finer features for functional microdevices}},
    year = {2001},
    journal = {Nature 2001 412:6848},
    author = {Kawata, S. and Sun, H. B. and Tanaka, T. and Takada, K.},
    number = {6848},
    month = {8},
    pages = {697--698},
    volume = {412},
    publisher = {Nature Publishing Group},
    url = {https://www.nature.com/articles/35089130},
    doi = {10.1038/35089130},
    issn = {1476-4687},
    pmid = {11507627}
}

@article{Nagy2019GenerationPower,
    title = {{Generation of three-cycle multi-millijoule laser pulses at 318 W average power}},
    year = {2019},
    journal = {Optica, Vol. 6, Issue 11, pp. 1423-1424},
    author = {Nagy, Tamas and H{\"{a}}drich, Steffen and H{\"{a}}drich, Steffen and Simon, Peter and Simon, Peter and Blumenstein, Andreas and Walther, Nico and Klas, Robert and Klas, Robert and Buldt, Joachim and Stark, Henning and Breitkopf, Sven and J{\'{o}}j{\'{a}}rt, Péter and Seres, Imre and V{\'{a}}rallyay, Zoltán and Eidam, Tino and Limpert, Jens and Limpert, Jens and Limpert, Jens and Limpert, Jens},
    number = {11},
    month = {11},
    pages = {1423--1424},
    volume = {6},
    publisher = {Optica Publishing Group},
    url = {https://opg.optica.org/viewmedia.cfm?uri=optica-6-11-1423&seq=0&html=true https://opg.optica.org/abstract.cfm?uri=optica-6-11-1423 https://opg.optica.org/optica/abstract.cfm?uri=optica-6-11-1423},
    doi = {10.1364/OPTICA.6.001423},
    issn = {2334-2536}
}

@article{Stearns1988GenerationPulses,
    title = {{Generation of ultrashort x-ray pulses}},
    year = {1988},
    journal = {Physical Review A},
    author = {Stearns, D. G. and Landen, O. L. and Campbell, E. M. and Scofield, J. H.},
    number = {5},
    pages = {1684--1690},
    volume = {37},
    doi = {10.1103/PhysRevA.37.1684},
    issn = {10502947}
}

@article{Vallieres2024HighAir,
    title = {{High Dose-Rate MeV Electron Beam from a Tightly-Focused Femtosecond IR Laser in Ambient Air}},
    year = {2024},
    journal = {Laser and Photonics Reviews},
    author = {Valli{\`{e}}res, Simon and Powell, Jeffrey and Connell, Tanner and Evans, Michael and Lytova, Marianna and Fillion-Gourdeau, François and Fourmaux, Sylvain and Payeur, Stéphane and Lassonde, Philippe and MacLean, Steve and L{\'{e}}gar{\'{e}}, François},
    number = {2},
    month = {2},
    pages = {2300078},
    volume = {18},
    publisher = {John Wiley and Sons Inc},
    url = {/doi/pdf/10.1002/lpor.202300078 https://onlinelibrary.wiley.com/doi/abs/10.1002/lpor.202300078 https://onlinelibrary.wiley.com/doi/10.1002/lpor.202300078},
    doi = {10.1002/LPOR.202300078;PAGEGROUP:STRING:PUBLICATION},
    issn = {18638899}
}

@article{Metzner2026InfluenceAblation,
    title = {{Influence of plasma, surface, and angle on interlinked X-ray emission dynamics in femtosecond burst pulse ablation}},
    year = {2026},
    journal = {Scientific Reports 2026 16:1},
    author = {Metzner, Daniel and Rebentrost, Philipp and Lickschat, Peter and Lampke, Thomas and Wei{\ss}mantel, Steffen},
    number = {1},
    month = {1},
    pages = {885-},
    volume = {16},
    publisher = {Nature Publishing Group},
    url = {https://www.nature.com/articles/s41598-025-34221-x},
    doi = {10.1038/s41598-025-34221-x},
    issn = {2045-2322},
    pmid = {41501315}
}

@article{Holland2022InfluenceProcessing,
    title = {{Influence of Pulse Duration on X-ray Emission during Industrial Ultrafast Laser Processing}},
    year = {2022},
    journal = {Materials 2022, Vol. 15, Page 2257},
    author = {Holland, Julian and Weber, Rudolf and Sailer, Marc and Graf, Thomas},
    number = {6},
    month = {3},
    pages = {2257},
    volume = {15},
    publisher = {Multidisciplinary Digital Publishing Institute},
    url = {https://www.mdpi.com/1996-1944/15/6/2257/htm https://www.mdpi.com/1996-1944/15/6/2257},
    doi = {10.3390/MA15062257},
    issn = {1996-1944}
}

@article{Hochstrasser2000IntenseAir,
    title = {{Intense terahertz pulses by four-wave rectification in air}},
    year = {2000},
    journal = {Optics Letters, Vol. 25, Issue 16, pp. 1210-1212},
    author = {Hochstrasser, R. M. and Cook, D. J.},
    number = {16},
    month = {8},
    pages = {1210--1212},
    volume = {25},
    publisher = {Optica Publishing Group},
    url = {https://opg.optica.org/viewmedia.cfm?uri=ol-25-16-1210&seq=0&html=true https://opg.optica.org/abstract.cfm?uri=ol-25-16-1210 https://opg.optica.org/ol/abstract.cfm?uri=ol-25-16-1210},
    doi = {10.1364/OL.25.001210},
    issn = {1539-4794},
    pmid = {18066171}
}

@article{Jung2025InterfaceSpectroscopy,
    title = {{Interface identification in micro-LED repair applications via depth profiling using femtosecond laser-induced breakdown spectroscopy}},
    year = {2025},
    journal = {Scientific Reports 2025 15:1},
    author = {Jung, Woonkyeong and Choi, Janghee and Jeon, Gookseon and Kim, Na Yoon and An, Jeongcheol and Jang, Inseok and Jeong, Sungho and Kim, Young Joo and Keum, Hohyun},
    number = {1},
    month = {11},
    pages = {40897-},
    volume = {15},
    publisher = {Nature Publishing Group},
    url = {https://www.nature.com/articles/s41598-025-24765-3},
    doi = {10.1038/s41598-025-24765-3},
    issn = {2045-2322},
    pmid = {41258172}
}

@book{Chen2016IntroductionFusion,
    title = {{Introduction to Plasma Physics and Controlled Fusion}},
    year = {2016},
    booktitle = {Introduction to Plasma Physics and Controlled Fusion},
    author = {Chen, Francis F.},
    publisher = {Springer International Publishing},
    doi = {10.1007/978-3-319-22309-4}
}

@article{Davies2009LaserRegime,
    title = {{Laser absorption by overdense plasmas in the relativistic regime}},
    year = {2009},
    journal = {Plasma Physics and Controlled Fusion},
    author = {Davies, J. R.},
    number = {1},
    month = {12},
    pages = {014006},
    volume = {51},
    publisher = {IOP Publishing},
    doi = {10.1088/0741-3335/51/1/014006},
    issn = {07413335}
}

@article{Tajima1979LaserAccelerator,
    title = {{Laser Electron Accelerator}},
    year = {1979},
    journal = {Physical Review Letters},
    author = {Tajima, T. and Dawson, J. M.},
    number = {4},
    month = {7},
    pages = {267},
    volume = {43},
    publisher = {American Physical Society},
    url = {https://journals.aps.org/prl/abstract/10.1103/PhysRevLett.43.267},
    doi = {10.1103/PhysRevLett.43.267},
    issn = {00319007}
}

@article{Zewail1988LaserFemtochemistry,
    title = {{Laser femtochemistry}},
    year = {1988},
    journal = {Science},
    author = {Zewail, Ahmed H.},
    number = {4886},
    month = {12},
    pages = {1645--1653},
    volume = {242},
    publisher = {American Association for the Advancement of Science},
    url = {/doi/pdf/10.1126/science.242.4886.1645?download=true},
    doi = {10.1126/SCIENCE.242.4886.1645;WEBSITE:WEBSITE:AAAS-SITE;JOURNAL:JOURNAL:SCIENCE;ISSUE:ISSUE:DOI},
    issn = {00368075}
}

@article{Miaja-Avila2015LaserSpectroscopy,
    title = {{Laser plasma x-ray source for ultrafast time-resolved x-ray absorption spectroscopy}},
    year = {2015},
    journal = {Structural Dynamics},
    author = {Miaja-Avila, L. and O'Neil, G. C. and Uhlig, J. and Cromer, C. L. and Dowell, M. L. and Jimenez, R. and Hoover, A. S. and Silverman, K. L. and Ullom, J. N.},
    number = {2},
    month = {3},
    volume = {2},
    publisher = {AAPM - American Association of Physicists in Medicine},
    doi = {10.1063/1.4913585},
    issn = {23297778}
}

@incollection{Kim1992Laser-ProducedMicroscopy,
    title = {{Laser-Produced Plasma as a Source for X-Ray Microscopy}},
    year = {1992},
    author = {Kim, H. and Yaakibi, B. and Soures, J. M. and Cheng, P. C.},
    pages = {47--53},
    publisher = {Springer, Berlin, Heidelberg},
    doi = {10.1007/978-3-540-46887-5{\_}8}
}

@article{Picksley2024MatchedAccelerators,
    title = {{Matched Guiding and Controlled Injection in Dark-Current-Free, 10-GeV-Class, Channel-Guided Laser-Plasma Accelerators}},
    year = {2024},
    journal = {Physical Review Letters},
    author = {Picksley, A. and Stackhouse, J. and Benedetti, C. and Nakamura, K. and Tsai, H. E. and Li, R. and Miao, B. and Shrock, J. E. and Rockafellow, E. and Milchberg, H. M. and Schroeder, C. B. and Van Tilborg, J. and Esarey, E. and Geddes, C. G.R. and Gonsalves, A. J.},
    number = {25},
    month = {12},
    pages = {255001},
    volume = {133},
    publisher = {American Physical Society},
    url = {https://journals.aps.org/prl/abstract/10.1103/PhysRevLett.133.255001},
    doi = {10.1103/PhysRevLett.133.255001},
    issn = {10797114},
    pmid = {39752715},
    arxivId = {2408.00740}
}

@article{Back2003Multi-keVPlasmas,
    title = {{Multi-keV x-ray conversion efficiency in laser-produced plasmas}},
    year = {2003},
    journal = {Physics of Plasmas},
    author = {Back, C. A. and Davis, J. and Grun, J. and Suter, L. J. and Landen, O. L. and Hsing, W. W. and Miller, M. C.},
    number = {5},
    month = {5},
    pages = {2047--2055},
    volume = {10},
    publisher = {AIP Publishing},
    doi = {10.1063/1.1566750},
    issn = {1070664X}
}

@article{Ferray1988Multiple-harmonicGases,
    title = {{Multiple-harmonic conversion of 1064 nm radiation in rare gases}},
    year = {1988},
    journal = {Journal of Physics B: Atomic, Molecular and Optical Physics},
    author = {Ferray, M. and L'Huillier, A. and Li, X. F. and Lompre, L. A. and Mainfray, G. and Manus, C.},
    number = {3},
    month = {2},
    pages = {L31},
    volume = {21},
    publisher = {IOP Publishing},
    url = {https://iopscience.iop.org/article/10.1088/0953-4075/21/3/001 https://iopscience.iop.org/article/10.1088/0953-4075/21/3/001/meta},
    doi = {10.1088/0953-4075/21/3/001},
    issn = {0953-4075}
}

@article{Udem2002OpticalMetrology,
    title = {{Optical frequency metrology}},
    year = {2002},
    journal = {Nature 2002 416:6877},
    author = {Udem, Th and Holzwarth, R. and H{\"{a}}nsch, T. W.},
    number = {6877},
    month = {3},
    pages = {233--237},
    volume = {416},
    publisher = {Nature Publishing Group},
    url = {https://www.nature.com/articles/416233a},
    isbn = {9,192,631,770},
    doi = {10.1038/416233a},
    issn = {1476-4687},
    pmid = {11894107}
}

@article{Steingrube2009PhaseCell,
    title = {{Phase matching of high-order harmonics in a semi-infinite gas cell}},
    year = {2009},
    journal = {Physical Review A},
    author = {Steingrube, Daniel S. and Vockerodt, Tobias and Schulz, Emilia and Morgner, Uwe and Kova{\v{c}}ev, Milutin},
    number = {4},
    month = {10},
    pages = {043819},
    volume = {80},
    publisher = {American Physical Society},
    url = {https://journals.aps.org/pra/abstract/10.1103/PhysRevA.80.043819},
    doi = {10.1103/PhysRevA.80.043819},
    issn = {10502947}
}

@article{Isinger2017PhotoionizationDomain,
    title = {{Photoionization in the time and frequency domain}},
    year = {2017},
    journal = {Science},
    author = {Isinger, M. and Squibb, R. J. and Busto, D. and Zhong, S. and Harth, A. and Kroon, D. and Nandi, S. and Arnold, C. L. and Miranda, M. and Dahlstr{\"{o}}m, J. M. and Lindroth, E. and Feifel, R. and Gisselbrecht, M. and L’Huillier, A.},
    number = {6365},
    month = {11},
    pages = {893--896},
    volume = {358},
    publisher = {American Association for the Advancement of Science},
    url = {/doi/pdf/10.1126/science.aao7043},
    doi = {10.1126/SCIENCE.AAO7043;SUBPAGE:STRING:ABSTRACT;ISSUE:ISSUE:DOI},
    issn = {10959203},
    pmid = {29097491},
    arxivId = {1709.01780}
}

@article{Rajhans2023Post-compressionRegime,
    title = {{Post-compression of multi-millijoule picosecond pulses to few-cycles approaching the terawatt regime}},
    year = {2023},
    journal = {Optics Letters, Vol. 48, Issue 18, pp. 4753-4756},
    author = {Rajhans, Supriya and Rajhans, Supriya and Escoto, Esmerando and Khodakovskiy, Nikita and Velpula, Praveen K. and Farace, Bonaventura and Grosse-Wortmann, Uwe and Shalloo, Rob J. and Arnold, Cord L. and Põder, Kristjan and Osterhoff, Jens and Leemans, Wim P. and Hartl, Ingmar and Heyl, Christoph M. and Heyl, Christoph M. and Heyl, Christoph M.},
    number = {18},
    month = {9},
    pages = {4753--4756},
    volume = {48},
    publisher = {Optica Publishing Group},
    url = {https://opg.optica.org/viewmedia.cfm?uri=ol-48-18-4753&seq=0&html=true https://opg.optica.org/abstract.cfm?uri=ol-48-18-4753 https://opg.optica.org/ol/abstract.cfm?uri=ol-48-18-4753},
    doi = {10.1364/OL.498042},
    issn = {1539-4794},
    pmid = {37707894},
    arxivId = {2306.09674}
}

@article{Dantus1987RealtimeReactions,
    title = {{Real‐time femtosecond probing of ‘‘transition states’’ in chemical reactions}},
    year = {1987},
    journal = {The Journal of Chemical Physics},
    author = {Dantus, Marcos and Rosker, Mark J. and Zewail, Ahmed H.},
    number = {4},
    month = {8},
    pages = {2395--2397},
    volume = {87},
    publisher = {AIP Publishing},
    url = {/aip/jcp/article/87/4/2395/220386/Real-time-femtosecond-probing-of-transition-states},
    doi = {10.1063/1.453122},
    issn = {0021-9606}
}

@article{Balmer1977ResonancePlasma,
    title = {{Resonance absorption of 1.06- {$\mu$}m laser radiation in laser-generated plasma}},
    year = {1977},
    journal = {Physical Review Letters},
    author = {Balmer, J. E. and Donaldson, T. P.},
    number = {17},
    month = {10},
    pages = {1084--1087},
    volume = {39},
    publisher = {American Physical Society},
    doi = {10.1103/PhysRevLett.39.1084},
    issn = {00319007}
}

@misc{Legall2021ReviewProcessing,
    title = {{Review of x-ray exposure and safety issues arising from ultra-short pulse laser material processing}},
    year = {2021},
    booktitle = {Journal of Radiological Protection},
    author = {Legall, Herbert and Bonse, Jörn and Kr{\"{u}}ger, Jörg},
    number = {1},
    month = {3},
    pages = {R28-R42},
    volume = {41},
    publisher = {IOP Publishing Ltd},
    doi = {10.1088/1361-6498/abcb16},
    issn = {13616498},
    pmid = {33202388}
}

@incollection{Bunte2004SecondaryEmission,
    title = {{Secondary Hazards: Particle and X-Ray Emission}},
    year = {2004},
    author = {Bunte, Jens and Barcikowski, Stephan and Puester, Thomas and Burmester, Tomas and Brose, Martin and Ludwig, Thomas},
    month = {9},
    pages = {309--321},
    publisher = {Springer, Berlin, Heidelberg},
    doi = {10.1007/978-3-540-39848-6{\_}20}
}

@article{Holland2024Self-ShieldingZone,
    title = {{Self-Shielding of X-ray Emission from Ultrafast Laser Processing Due to Geometrical Changes of the Interaction Zone}},
    year = {2024},
    journal = {Materials},
    author = {Holland, Julian and Hagenlocher, Christian and Weber, Rudolf and Graf, Thomas},
    number = {5},
    month = {3},
    pages = {1109},
    volume = {17},
    publisher = {Multidisciplinary Digital Publishing Institute (MDPI)},
    doi = {10.3390/ma17051109},
    issn = {19961944}
}

@book{Gibbon2005ShortIntroduction,
    title = {{Short pulse laser interactions with matter: An introduction}},
    year = {2005},
    booktitle = {Short Pulse Laser Interactions with Matter: An Introduction},
    author = {Gibbon, Paul},
    month = {1},
    pages = {1--312},
    publisher = {Imperial College Press},
    isbn = {9781860949340},
    doi = {10.1142/P116}
}

@article{Kuhlke1987SoftPlasmas,
    title = {{Soft x-ray emission from subpicosecond laser-produced plasmas}},
    year = {1987},
    journal = {Applied Physics Letters},
    author = {K{\"{u}}hlke, D. and Herpers, U. and Von Der Linde, D.},
    number = {25},
    pages = {1785--1787},
    volume = {50},
    doi = {10.1063/1.97696},
    issn = {00036951}
}

@article{Murray1997SpatialIssues,
    title = {{Spatial filter issues}},
    year = {1997},
    journal = {https://doi.org/10.1117/12.294305},
    author = {Murray, James E. and Estabrook, Kent G. and Milam, David and Sell, Walter D. and Wonterghem, Bruno M. Van and Feit, Michael D. and Rubenchik, Alexander M.},
    month = {12},
    pages = {207--212},
    volume = {3047},
    publisher = {SPIE},
    url = {https://www.spiedigitallibrary.org/conference-proceedings-of-spie/3047/0000/Spatial-filter-issues/10.1117/12.294305.full https://www.spiedigitallibrary.org/conference-proceedings-of-spie/3047/0000/Spatial-filter-issues/10.1117/12.294305.short},
    doi = {10.1117/12.294305}
}

@article{McIntyre1987StudiesGases,
    title = {{Studies of multiphoton production of vacuum-ultraviolet radiation in the rare gases}},
    year = {1987},
    journal = {JOSA B, Vol. 4, Issue 4, pp. 595-601},
    author = {McIntyre, I. A. and Jara, H. and Johann, U. and Luk, T. S. and Rhodes, C. K. and McPherson, A. and Gibson, G. and Boyer, K.},
    number = {4},
    month = {4},
    pages = {595--601},
    volume = {4},
    publisher = {Optica Publishing Group},
    url = {https://opg.optica.org/viewmedia.cfm?uri=josab-4-4-595&seq=0&html=true https://opg.optica.org/abstract.cfm?uri=josab-4-4-595 https://opg.optica.org/josab/abstract.cfm?uri=josab-4-4-595},
    doi = {10.1364/JOSAB.4.000595},
    issn = {1520-8540}
}

@article{Schille2021StudyProcessing,
    title = {{Study on x-ray emission using ultrashort pulsed lasers in materials processing}},
    year = {2021},
    journal = {Materials},
    author = {Schille, Joerg and Kraft, Sebastian and Pflug, Theo and Scholz, Christian and Clair, Maurice and Horn, Alexander and Loeschner, Udo},
    number = {16},
    month = {8},
    volume = {14},
    publisher = {MDPI AG},
    doi = {10.3390/ma14164537},
    issn = {19961944}
}

@article{Gardner2017SubwavelengthSource,
    title = {{Subwavelength coherent imaging of periodic samples using a 13.5 nm tabletop high-harmonic light source}},
    year = {2017},
    journal = {Nature Photonics 2017 11:4},
    author = {Gardner, Dennis F. and Tanksalvala, Michael and Shanblatt, Elisabeth R. and Zhang, Xiaoshi and Galloway, Benjamin R. and Porter, Christina L. and Karl, Robert and Bevis, Charles and Adams, Daniel E. and Kapteyn, Henry C. and Murnane, Margaret M. and Mancini, Giulia F.},
    number = {4},
    month = {3},
    pages = {259--263},
    volume = {11},
    publisher = {Nature Publishing Group},
    url = {https://www.nature.com/articles/nphoton.2017.33},
    doi = {10.1038/nphoton.2017.33},
    issn = {1749-4893}
}

@article{Appi2021SynchronizedStudies,
    title = {{Synchronized beamline at FLASH2 based on high-order harmonic generation for two-color dynamics studies}},
    year = {2021},
    journal = {Cite as: Rev. Sci. Instrum},
    author = {Appi, E and Papadopoulou, C C and Mapa, J L and Jusko, C and Mosel, P and Schoenberg, A and Stock, J and Feigl, T and Ali{\v{s}}auskas, S and Lang, T and Heyl, C M and Manschwetus, B and Brachmanski, M and Braune, M and Lindenblatt, H and Trost, F and Meister, S and Schoch, P and Trabattoni, A and Calegari, F and Treusch, R and Moshammer, R and Hartl, I and Morgner, U and Kovacev, M},
    pages = {123004},
    volume = {92},
    url = {http://creativecommons.org/licenses/by/4.0/},
    doi = {10.1063/5.0063225}
}

@article{Gardner2014TabletopPtychography,
    title = {{Tabletop nanometer extreme ultraviolet imaging in an extended reflection mode using coherent Fresnel ptychography}},
    year = {2014},
    journal = {Optica, Vol. 1, Issue 1, pp. 39-44},
    author = {Gardner, Dennis F. and Kapteyn, Henry C. and Seaberg, Matthew D. and Shanblatt, Elisabeth R. and Murnane, Margaret M. and Adams, Daniel E. and Zhang, Bosheng},
    number = {1},
    month = {7},
    pages = {39--44},
    volume = {1},
    publisher = {Optica Publishing Group},
    url = {https://opg.optica.org/viewmedia.cfm?uri=optica-1-1-39&seq=0&html=true https://opg.optica.org/abstract.cfm?uri=optica-1-1-39 https://opg.optica.org/optica/abstract.cfm?uri=optica-1-1-39},
    doi = {10.1364/OPTICA.1.000039},
    issn = {2334-2536},
    arxivId = {1312.2049}
}

@article{Weber1988TemporalCavities,
    title = {{Temporal and spectral characteristics of soft x radiation from laser-irradiated cylindrical cavities}},
    year = {1988},
    journal = {Applied Physics Letters},
    author = {Weber, R. and Cunningham, P. F. and Balmer, J. E.},
    number = {26},
    month = {12},
    pages = {2596--2598},
    volume = {53},
    publisher = {AIP Publishing},
    doi = {10.1063/1.100190},
    issn = {00036951}
}

@article{Jepsen2011TerahertzApplications,
    title = {{Terahertz spectroscopy and imaging - Modern techniques and applications}},
    year = {2011},
    journal = {Laser and Photonics Reviews},
    author = {Jepsen, P. U. and Cooke, D. G. and Koch, M.},
    number = {1},
    month = {1},
    pages = {124--166},
    volume = {5},
    publisher = {John Wiley {\&} Sons, Ltd},
    url = {/doi/pdf/10.1002/lpor.201000011 https://onlinelibrary.wiley.com/doi/abs/10.1002/lpor.201000011 https://onlinelibrary.wiley.com/doi/10.1002/lpor.201000011},
    doi = {10.1002/LPOR.201000011;PAGE:STRING:ARTICLE/CHAPTER},
    issn = {18638880}
}

@article{Legall2019TheMachining,
    title = {{The influence of processing parameters on X-ray emission during ultra-short pulse laser machining}},
    year = {2019},
    journal = {Applied Physics A: Materials Science and Processing},
    author = {Legall, Herbert and Schwanke, Christoph and Bonse, Jörn and Kr{\"{u}}ger, Jörg},
    number = {8},
    month = {8},
    volume = {125},
    publisher = {Springer Verlag},
    doi = {10.1007/s00339-019-2827-y},
    issn = {14320630}
}

@book{Eliezer2002ThePlasmas,
    title = {{The interaction of high-power lasers with plasmas}},
    year = {2002},
    author = {Eliezer, Shalom.},
    pages = {323},
    publisher = {Institute of Physics Pub.},
    isbn = {978-0-7503-0747-5}
}

@article{Corkum1988ThermalExcitation,
    title = {{Thermal response of metals to ultrashort-pulse laser excitation}},
    year = {1988},
    journal = {Physical Review Letters},
    author = {Corkum, P. B. and Brunel, F. and Sherman, N. K. and Srinivasan-Rao, T.},
    number = {25},
    month = {12},
    pages = {2886--2889},
    volume = {61},
    publisher = {American Physical Society},
    doi = {10.1103/PhysRevLett.61.2886},
    issn = {00319007},
    pmid = {10039253}
}

@article{Malinauskas2016UltrafastIndustry,
    title = {{Ultrafast laser processing of materials: from science to industry}},
    year = {2016},
    journal = {Light: Science {\&} Applications 2016 5:8},
    author = {Malinauskas, Mangirdas and {\v{Z}}ukauskas, Albertas and Hasegawa, Satoshi and Hayasaki, Yoshio and Mizeikis, Vygantas and Buividas, Ričardas and Juodkazis, Saulius},
    number = {8},
    month = {3},
    pages = {e16133-e16133},
    volume = {5},
    publisher = {Nature Publishing Group},
    url = {https://www.nature.com/articles/lsa2016133},
    doi = {10.1038/lsa.2016.133},
    issn = {2047-7538}
}

@article{Murnane1991UltrafastPlasmas,
    title = {{Ultrafast X-ray pulses from laser-produced plasmas}},
    year = {1991},
    journal = {Science},
    author = {Murnane, Margaret M. and Kapteyn, Henry C. and Rosen, Mordecai D. and Falcone, Roger W.},
    number = {4993},
    month = {2},
    pages = {531--536},
    volume = {251},
    publisher = {American Association for the Advancement of Science},
    doi = {10.1126/science.251.4993.531},
    issn = {00368075}
}

@article{Davis1996WritingLaser,
    title = {{Writing waveguides in glass with a femtosecond laser}},
    year = {1996},
    journal = {Optics Letters, Vol. 21, Issue 21, pp. 1729-1731},
    author = {Davis, K. M. and Miura, K. and Hirao, K. and Sugimoto, N.},
    number = {21},
    month = {11},
    pages = {1729--1731},
    volume = {21},
    publisher = {Optica Publishing Group},
    url = {https://opg.optica.org/viewmedia.cfm?uri=ol-21-21-1729&seq=0&html=true https://opg.optica.org/abstract.cfm?uri=ol-21-21-1729 https://opg.optica.org/ol/abstract.cfm?uri=ol-21-21-1729},
    doi = {10.1364/OL.21.001729},
    issn = {1539-4794},
    pmid = {19881782}
}

@article{Mosel2021X-rayDetector,
    title = {{X-ray dose rate and spectral measurements during ultrafast laser machining using a calibrated (High-sensitivity) novel x-ray detector}},
    year = {2021},
    journal = {Materials},
    author = {Mosel, Philip and Sankar, Pranitha and D{\"{u}}sing, Jan Friedrich and Dittmar, Günter and P{\"{u}}ster, Thomas and J{\"{a}}schke, Peter and Vahlbruch, Jan Willem and Morgner, Uwe and Kovacev, Milutin},
    number = {16},
    month = {8},
    volume = {14},
    publisher = {MDPI AG},
    doi = {10.3390/ma14164397},
    issn = {19961944}
}

@article{Legall2018X-rayProcessing,
    title = {{X-ray emission as a potential hazard during ultrashort pulse laser material processing}},
    year = {2018},
    journal = {Applied Physics A: Materials Science and Processing},
    author = {Legall, Herbert and Schwanke, Christoph and Pentzien, Simone and Dittmar, Günter and Bonse, Jörn and Kr{\"{u}}ger, Jörg},
    number = {6},
    month = {6},
    pages = {1--8},
    volume = {124},
    publisher = {Springer Verlag},
    doi = {10.1007/s00339-018-1828-6},
    issn = {14320630}
}

@inproceedings{Weber2022X-rayProblem,
    title = {{X-ray emission during materials processing with ultrashort laser pulses - A serious problem?}},
    year = {2022},
    booktitle = {Procedia CIRP},
    author = {Weber, Rudolf and Graf, Thomas},
    month = {1},
    pages = {844--849},
    volume = {111},
    publisher = {Elsevier B.V.},
    doi = {10.1016/j.procir.2022.08.095},
    issn = {22128271}
}

@article{Thogersen2001X-rayMicromachining,
    title = {{X-ray emission from femtosecond laser micromachining}},
    year = {2001},
    journal = {Applied Physics A: Materials Science and Processing},
    author = {Thogersen, J. and Borowiec, A. and Haugen, H. K. and McNeill, F. E. and Stronach, I. M.},
    number = {3},
    pages = {361--363},
    volume = {73},
    publisher = {Springer Verlag},
    url = {https://link.springer.com/article/10.1007/s003390100741},
    doi = {10.1007/S003390100741/METRICS},
    issn = {09478396}
}

@article{Giulietti1998X-rayPlasmas,
    title = {{X-ray emission from laser-produced plasmas}},
    year = {1998},
    journal = {Rivista del Nuovo Cimento della Societa Italiana di Fisica},
    author = {Giulietti, Danilo and Gizzi, Leonida A.},
    number = {10},
    pages = {1--93},
    volume = {21},
    publisher = {Editrice Compositori s.r.l.},
    doi = {10.1007/BF02874624},
    issn = {0393697X}
}

@article{Behrens2019X-RAYLASERS,
    title = {{X-RAY EMISSION FROM MATERIALS PROCESSING LASERS}},
    year = {2019},
    journal = {Radiation Protection Dosimetry},
    author = {Behrens, R and Pullner, B and Reginatto, M},
    number = {3},
    month = {5},
    pages = {361--374},
    volume = {183},
    publisher = {Oxford University Press},
    url = {https://academic.oup.com/rpd/article/183/3/361/5090842},
    doi = {10.1093/rpd/ncy126},
    issn = {0144-8420}
}

@article{Stolzenberg2021X-raySetting,
    title = {{X-ray emission hazards from ultrashort pulsed laser material processing in an industrial setting}},
    year = {2021},
    journal = {Materials},
    author = {Stolzenberg, Ulf and Schmitt Rahner, Mayka and Pullner, Björn and Legall, Herbert and Bonse, Jörn and Kluge, Michael and Ortner, Andreas and Hoppe, Bernd and Kr{\"{u}}ger, Jörg},
    number = {23},
    month = {12},
    pages = {7163},
    volume = {14},
    publisher = {MDPI},
    doi = {10.3390/ma14237163},
    issn = {19961944}
}

@article{Metzner2021X-rayBursts,
    title = {{X-ray generation by laser ablation using MHz to GHz pulse bursts}},
    year = {2021},
    journal = {Journal of Laser Applications},
    author = {Metzner, Daniel and Olbrich, Markus and Lickschat, Peter and Horn, Alexander and Wei{\ss}mantel, Steffen},
    number = {3},
    month = {8},
    volume = {33},
    publisher = {Laser Institute of America},
    doi = {10.2351/7.0000403},
    issn = {1042-346X}
}

@article{Legall2020X-rayProcessing,
    title = {{X-ray radiation protection aspects during ultrashort laser processing}},
    year = {2020},
    journal = {Journal of Laser Applications},
    author = {Legall, Herbert and Schwanke, Christoph and Bonse, Jörn and Kr{\"{u}}ger, Jörg},
    number = {2},
    month = {5},
    volume = {32},
    publisher = {Laser Institute of America},
    doi = {10.2351/1.5134778},
    issn = {1042-346X}
}

@article{Freitag2020XRayEnvironment,
    title = {{X‐Ray Protection in an Industrial Production Environment}},
    year = {2020},
    journal = {PhotonicsViews},
    author = {Freitag, Christian and Giedl‐Wagner, Roswitha},
    number = {3},
    month = {6},
    pages = {37--41},
    volume = {17},
    publisher = {Wiley},
    doi = {10.1002/phvs.202000020},
    issn = {2626-1294}
}

@article{Neuenschwander2010,
   author = {Beat Neuenschwander and Guido F. Bucher and Christian Nussbaum and Benjamin Joss and Martin Muralt and Urs W. Hunziker and Peter Schuetz},
   doi = {10.1117/12.846521},
   isbn = {9780819479808},
   issn = {0277786X},
   journal = {https://doi.org/10.1117/12.846521},
   month = {2},
   pages = {99-112},
   publisher = {SPIE},
   title = {Processing of metals and dielectric materials with ps-laserpulses: results, strategies, limitations and needs},
   volume = {7584},
   url = {https://www.spiedigitallibrary.org/conference-proceedings-of-spie/7584/75840R/Processing-of-metals-and-dielectric-materials-with-ps-laserpulses/10.1117/12.846521.full https://www.spiedigitallibrary.org/conference-proceedings-of-spie/7584/75840R/Processing-of-metals-and-dielectric-materials-with-ps-laserpulses/10.1117/12.846521.short},
   year = {2010}
}

@article{Lauer2014,
   author = {Benjamin Lauer and Beat Jäggi and Beat Neuenschwander},
   doi = {10.1016/J.PHPRO.2014.08.116},
   issn = {1875-3892},
   issue = {C},
   journal = {Physics Procedia},
   month = {1},
   pages = {963-972},
   publisher = {Elsevier},
   title = {Influence of the Pulse Duration onto the Material Removal Rate and Machining Quality for Different Types of Steel},
   volume = {56},
   year = {2014}
}
\end{document}